\input graphicx

\input harvmac
\input amssym
\input epsf

\def\unit{\relax{\rm 1\kern-.26em I}}
\def\nada{\relax{\rm 0\kern-.30em l}}
\def\tilde{\widetilde}

\def\det{{\rm det}}

\noblackbox
\def\IL{\relax{\rm I\kern-.18em L}}
\def\IH{\relax{\rm I\kern-.18em H}}
\def\IR{\relax{\rm I\kern-.18em R}}
\def\IC{\relax\hbox{$\inbar\kern-.3em{\rm C}$}}
\def\IZ{\relax\ifmmode\mathchoice
{\hbox{\cmss Z\kern-.4em Z}}{\hbox{\cmss Z\kern-.4em Z}}
{\lower.9pt\hbox{\cmsss Z\kern-.4em Z}} {\lower1.2pt\hbox{\cmsss
Z\kern-.4em Z}}\else{\cmss Z\kern-.4em Z}\fi}
\def\CM {{\cal M}}

\def\CL {{\cal L}}

\def\CO {{\cal O}}

\def\CM {{\cal M}}

\def\CO {{\cal O}}

\def\det{{\rm det}}
\def\Tr{{\rm Tr}}

\font\manual=manfnt \def\dbend{\lower3.5pt\hbox{\manual\char127}}

\def\IZ{\relax\ifmmode\mathchoice
{\hbox{\cmss Z\kern-.4em Z}}{\hbox{\cmss Z\kern-.4em Z}}
{\lower.9pt\hbox{\cmsss Z\kern-.4em Z}} {\lower1.2pt\hbox{\cmsss
Z\kern-.4em Z}}\else{\cmss Z\kern-.4em Z}\fi}

\def\lfm#1{\medskip\noindent\item{#1}}

\def\bar{\overline}

\def\rt2{\sqrt{2}}
\def\irt2{{1\over\sqrt{2}}}

\def\slashchar#1{\setbox0=\hbox{$#1$}           
\dimen0=\wd0                                 
\setbox1=\hbox{/} \dimen1=\wd1               
\ifdim\dimen0>\dimen1                        
   \rlap{\hbox to \dimen0{\hfil/\hfil}}      
   #1                                        
\else                                        
   \rlap{\hbox to \dimen1{\hfil$#1$\hfil}}   
   /                                         
\fi}

\def\foursqr#1#2{{\vcenter{\vbox{
 \hrule height.#2pt
 \hbox{\vrule width.#2pt height#1pt \kern#1pt
 \vrule width.#2pt}
 \hrule height.#2pt
 \hrule height.#2pt
 \hbox{\vrule width.#2pt height#1pt \kern#1pt
 \vrule width.#2pt}
 \hrule height.#2pt
     \hrule height.#2pt
 \hbox{\vrule width.#2pt height#1pt \kern#1pt
 \vrule width.#2pt}
 \hrule height.#2pt
     \hrule height.#2pt
 \hbox{\vrule width.#2pt height#1pt \kern#1pt
 \vrule width.#2pt}
 \hrule height.#2pt}}}}
\def\psqr#1#2{{\vcenter{\vbox{\hrule height.#2pt
 \hbox{\vrule width.#2pt height#1pt \kern#1pt
 \vrule width.#2pt}
 \hrule height.#2pt \hrule height.#2pt
 \hbox{\vrule width.#2pt height#1pt \kern#1pt
 \vrule width.#2pt}
 \hrule height.#2pt}}}}
\def\sqr#1#2{{\vcenter{\vbox{\hrule height.#2pt
 \hbox{\vrule width.#2pt height#1pt \kern#1pt
 \vrule width.#2pt}
 \hrule height.#2pt}}}}

\def\figin{\epsfcheck\figin}\def\figins{\epsfcheck\figins}
\def\epsfcheck{\ifx\epsfbox\UnDeFiNeD
\message{(NO epsf.tex, FIGURES WILL BE IGNORED)}
\gdef\figin##1{\vskip2in}\gdef\figins##1{\hskip.5in}
\else\message{(FIGURES WILL BE INCLUDED)}%
\gdef\figin##1{##1}\gdef\figins##1{##1}\fi}
\def\DefWarn#1{}
\def\figinsert{\goodbreak\midinsert}
\def\ifig#1#2#3{\DefWarn#1\xdef#1{fig.~\the\figno}
\writedef{#1\leftbracket fig.\noexpand~\the\figno}%
\figinsert\figin{\centerline{#3}}\medskip\centerline{\vbox{\baselineskip12pt
\advance\hsize by -1truein\noindent\footnotefont{\bf
Fig.~\the\figno:\ } \it#2}}
\bigskip\endinsert\global\advance\figno by1}

\lref\RayWK{
  S.~Ray,
  ``Some properties of meta-stable supersymmetry-breaking vacua in Wess-Zumino
  models,''
  Phys.\ Lett.\  B {\bf 642}, 137 (2006)
  [arXiv:hep-th/0607172].
}

\lref\CheungES{
  C.~Cheung, A.~L.~Fitzpatrick and D.~Shih,
  ``(Extra)Ordinary Gauge Mediation,''
  JHEP {\bf 0807}, 054 (2008)
  [arXiv:0710.3585 [hep-ph]].
}

\lref\DimopoulosGY{
  S.~Dimopoulos, G.~F.~Giudice and A.~Pomarol,
  ``Dark matter in theories of gauge-mediated supersymmetry breaking,''
  Phys.\ Lett.\  B {\bf 389}, 37 (1996)
  [arXiv:hep-ph/9607225].
}

\lref\MeadeWD{
  P.~Meade, N.~Seiberg and D.~Shih,
  ``General Gauge Mediation,''
  arXiv:0801.3278 [hep-ph].
}

\lref\KomargodskiAX{
  Z.~Komargodski and N.~Seiberg,
  ``mu and General Gauge Mediation,''
  arXiv:0812.3900 [hep-ph].
}

\lref\ThomasWY{
  S.~D.~Thomas and J.~D.~Wells,
  ``Phenomenology of Massive Vectorlike Doublet Leptons,''
  Phys.\ Rev.\ Lett.\  {\bf 81}, 34 (1998)
  [arXiv:hep-ph/9804359].
}

\lref\EssigAZ{
  R.~Essig,
  ``Direct Detection of Non-Chiral Dark Matter,''
  Phys.\ Rev.\  D {\bf 78}, 015004 (2008)
  [arXiv:0710.1668 [hep-ph]].
}
\lref\MahbubaniPT{
  R.~Mahbubani and L.~Senatore,
  ``The minimal model for dark matter and unification,''
  Phys.\ Rev.\  D {\bf 73}, 043510 (2006)
  [arXiv:hep-ph/0510064].
}

\lref\GiudiceBP{
  G.~F.~Giudice and R.~Rattazzi,
  ``Theories with gauge-mediated supersymmetry breaking,''
  Phys.\ Rept.\  {\bf 322}, 419 (1999)
  [arXiv:hep-ph/9801271].
}

\lref\IntriligatorDD{
  K.~A.~Intriligator, N.~Seiberg and D.~Shih,
  ``Dynamical SUSY breaking in meta-stable vacua,''
  JHEP {\bf 0604}, 021 (2006)
  [arXiv:hep-th/0602239].
}

\lref\IY{
  K.~I.~Izawa and T.~Yanagida,
  ``Dynamical Supersymmetry Breaking in Vector-like Gauge Theories,''
  Prog.\ Theor.\ Phys.\  {\bf 95}, 829 (1996)
  [arXiv:hep-th/9602180].
}

\lref\CheungES{
  C.~Cheung, A.~L.~Fitzpatrick and D.~Shih,
  ``(Extra)Ordinary Gauge Mediation,''
  JHEP {\bf 0807}, 054 (2008)
  [arXiv:0710.3585 [hep-ph]].
}

\lref\IntriligatorPU{
  K.~A.~Intriligator and S.~D.~Thomas,
  ``Dynamical Supersymmetry Breaking on Quantum Moduli Spaces,''
  Nucl.\ Phys.\  B {\bf 473}, 121 (1996)
  [arXiv:hep-th/9603158].
}

\lref\DineGM{
  M.~Dine, J.~L.~Feng and E.~Silverstein,
  ``Retrofitting O'Raifeartaigh models with dynamical scales,''
  Phys.\ Rev.\  D {\bf 74}, 095012 (2006)
  [arXiv:hep-th/0608159].
}

\lref\HanWN{
  T.~Han and R.~Hempfling,
  ``Messenger sneutrinos as cold dark matter,''
  Phys.\ Lett.\  B {\bf 415}, 161 (1997)
  [arXiv:hep-ph/9708264].
}

\lref\NelsonNF{
  A.~E.~Nelson and N.~Seiberg,
  ``R symmetry breaking versus supersymmetry breaking,''
  Nucl.\ Phys.\  B {\bf 416}, 46 (1994)
  [arXiv:hep-ph/9309299].
}

\lref\ShihAV{
  D.~Shih,
  ``Spontaneous R-symmetry breaking in O'Raifeartaigh models,''
  JHEP {\bf 0802}, 091 (2008)
  [arXiv:hep-th/0703196].
}

\lref\KomargodskiJF{
  Z.~Komargodski and D.~Shih,
  ``Notes on SUSY and R-Symmetry Breaking in Wess-Zumino Models,''
  JHEP {\bf 0904}, 093 (2009)
  [arXiv:0902.0030 [hep-th]].
}

\lref\TuckerSmithHY{
  D.~Tucker-Smith and N.~Weiner,
  ``Inelastic dark matter,''
  Phys.\ Rev.\  D {\bf 64}, 043502 (2001)
  [arXiv:hep-ph/0101138].
}

\lref\TuckerSmithJV{
  D.~Tucker-Smith and N.~Weiner,
  ``The status of inelastic dark matter,''
  Phys.\ Rev.\  D {\bf 72}, 063509 (2005)
  [arXiv:hep-ph/0402065].
}

\lref\ChangGD{
  S.~Chang, G.~D.~Kribs, D.~Tucker-Smith and N.~Weiner,
  ``Inelastic Dark Matter in Light of DAMA/LIBRA,''
  Phys.\ Rev.\  D {\bf 79}, 043513 (2009)
  [arXiv:0807.2250 [hep-ph]].
}

\lref\NussinovFT{
  S.~Nussinov, L.~T.~Wang and I.~Yavin,
  ``Capture of Inelastic Dark Matter in the Sun,''
  arXiv:0905.1333 [hep-ph].
}
\lref\MenonQJ{
  A.~Menon, R.~Morris, A.~Pierce and N.~Weiner,
  ``Capture and Indirect Detection of Inelastic Dark Matter,''
  arXiv:0905.1847 [hep-ph].
}

\lref\CirelliUQ{
  M.~Cirelli, N.~Fornengo and A.~Strumia,
  Nucl.\ Phys.\  B {\bf 753}, 178 (2006)
  [arXiv:hep-ph/0512090].
}

\lref\ArkaniHamedQP{
  N.~Arkani-Hamed and N.~Weiner,
  ``LHC Signals for a SuperUnified Theory of Dark Matter,''
  JHEP {\bf 0812}, 104 (2008)
  [arXiv:0810.0714 [hep-ph]].
}

\lref\ArvanitakiNQ{
  A.~Arvanitaki, S.~Dimopoulos, A.~Pierce, S.~Rajendran and J.~G.~Wacker,
  ``Stopping gluinos,''
  Phys.\ Rev.\  D {\bf 76}, 055007 (2007)
  [arXiv:hep-ph/0506242].
}

\lref\PospelovMP{
  M.~Pospelov, A.~Ritz and M.~B.~Voloshin,
  ``Secluded WIMP Dark Matter,''
  Phys.\ Lett.\  B {\bf 662}, 53 (2008)
  [arXiv:0711.4866 [hep-ph]].
}

\lref\MeadeIU{
  P.~Meade, M.~Papucci, A.~Strumia and T.~Volansky,
  ``Dark Matter Interpretations of the Electron/Positron Excesses after
  FERMI,''
  arXiv:0905.0480 [hep-ph].
}

\lref\ArvanitakiHQ{
  A.~Arvanitaki, S.~Dimopoulos, S.~Dubovsky, P.~W.~Graham, R.~Harnik and S.~Rajendran,
  ``Astrophysical Probes of Unification,''
  arXiv:0812.2075 [hep-ph].
}

\lref\HooperZM{
  D.~Hooper,
  ``TASI 2008 Lectures on Dark Matter,''
  arXiv:0901.4090 [hep-ph].
}

\lref\DineAG{
  M.~Dine, A.~E.~Nelson, Y.~Nir and Y.~Shirman,
  ``New tools for low-energy dynamical supersymmetry breaking,''
  Phys.\ Rev.\  D {\bf 53}, 2658 (1996)
  [arXiv:hep-ph/9507378].
}

\lref\DimopoulosGY{
  S.~Dimopoulos, G.~F.~Giudice and A.~Pomarol,
  ``Dark matter in theories of gauge-mediated supersymmetry breaking,''
  Phys.\ Lett.\  B {\bf 389}, 37 (1996)
  [arXiv:hep-ph/9607225].
}

\lref\ArkaniHamedQN{
  N.~Arkani-Hamed, D.~P.~Finkbeiner, T.~R.~Slatyer and N.~Weiner,
  ``A Theory of Dark Matter,''
  Phys.\ Rev.\  D {\bf 79}, 015014 (2009)
  [arXiv:0810.0713 [hep-ph]].
}

\lref\PDG{
 C.~Amsler et al.\ (Particle Data Group), Physics Letters B{\bf 667}, 1 (2008) 
}

\lref\HoldomAG{
  B.~Holdom,
  ``Two U(1)'S And Epsilon Charge Shifts,''
  Phys.\ Lett.\  B {\bf 166}, 196 (1986).
}

\lref\AffleckXZ{
  I.~Affleck, M.~Dine and N.~Seiberg,
  ``Dynamical Supersymmetry Breaking In Four-Dimensions And Its
  Phenomenological Implications,''
  Nucl.\ Phys.\  B {\bf 256}, 557 (1985).
}

\lref\KitanoXG{
  R.~Kitano, H.~Ooguri and Y.~Ookouchi,
  ``Direct mediation of meta-stable supersymmetry breaking,''
  Phys.\ Rev.\  D {\bf 75}, 045022 (2007)
  [arXiv:hep-ph/0612139].
}

\lref\HabaRJ{
  N.~Haba and N.~Maru,
  ``A Simple Model of Direct Gauge Mediation of Metastable Supersymmetry
  Breaking,''
  Phys.\ Rev.\  D {\bf 76}, 115019 (2007)
  [arXiv:0709.2945 [hep-ph]].
}

\lref\ZurZG{
  B.~K.~Zur, L.~Mazzucato and Y.~Oz,
  ``Direct Mediation and a Visible Metastable Supersymmetry Breaking Sector,''
  JHEP {\bf 0810}, 099 (2008)
  [arXiv:0807.4543 [hep-ph]].
}


\newbox\tmpbox\setbox\tmpbox\hbox{\abstractfont }
\Title{\vbox{\baselineskip12pt }} {\vbox{\centerline{ Pseudomoduli Dark Matter}}}
\smallskip
\centerline{David Shih}
\smallskip
\bigskip
\centerline{{\it School
of Natural Sciences, Institute for Advanced Study, Princeton, NJ
08540 USA}} \vskip 1cm

\noindent  We point out that pseudomoduli -- tree-level flat directions that often accompany dynamical supersymmetry breaking -- can be natural candidates for TeV-scale dark matter in models of gauge mediation. The idea is general and can be applied to different dark matter scenarios, including (but not limited to) those of potential relevance to recent cosmic ray anomalies. We describe the requirements for a viable model of pseudomoduli dark matter, and we analyze two example models to illustrate the general mechanism -- one where the pseudomoduli carry Higgsino-like quantum numbers, and another where they are SM singlets but are charged under a hidden-sector $U(1)'$ gauge group.

\bigskip

\Date{June 2009}

\newsec{Introduction}

By now, numerous astronomical observations and experiments have firmly established the existence of dark matter (for a recent review with references, see \HooperZM), with a cosmological abundance measured to be
\eqn\relicdens{
\Omega_{DM}h^2\approx 0.1
}
In the standard paradigm, the dark matter is a stable particle (a ``WIMP") which can annihilate to lighter Standard Model (SM) states. These annihilations maintain the dark matter in thermal equilibrium with the SM in the early universe. Gradually, as the universe cools and expands, the number density of dark matter decreases to the point that the dark matter ceases to annihilate efficiently (``freezes out"). This leaves behind a relic abundance which depends on the annihilation rate via the following well-known, approximate formula:
\eqn\relicdensii{
\Omega_{DM}h^2 \approx 0.1\times \left({3\times 10^{-26}\,\, {\rm cm}^3{\rm s}^{-1}
\over \langle\sigma v\rangle}\right)
}
For two-body tree-level annihilation to lighter states via some interaction with coupling constant $g$, one typically has (up to order-one factors)
 \eqn\xsecreqii{
 \langle\sigma v\rangle\sim {\alpha_g^2\over M_{DM}^2} \sim \alpha_g^2\times \left({1\,\,{\rm TeV}\over M_{DM}}\right)^{2} \times 10^{-23}\,\,{\rm cm}^3{\rm s}^{-1}
 }
where $\alpha_g\equiv {g^2/ 4\pi}$. So for $g\sim 1$ and $M_{DM}\sim 1$ TeV, one obtains a relic density with the right order of magnitude. 

The surprising appearance of the TeV scale strongly hints at a common origin for the dark matter mass and electroweak symmetry breaking. And since supersymmetry (SUSY) is a compelling explanation for the origin of the weak scale, it is commonly supposed that dark matter and SUSY must be intimately linked. This is indeed the case in scenarios of high-scale SUSY breaking, such as ``gravity mediation," where the lightest neutralino in the MSSM can be absolutely stable and can be a viable dark matter candidate. However, models of gravity mediation inevitably suffer from the SUSY flavor problem, whereby uncontrolled and uncalculable Planck-suppressed operators give rise to unobserved (and highly constrained) flavor-changing processes. 

For this reason, much theoretical effort has been devoted to studying alternative scenarios where SUSY breaking occurs and is mediated at lower scales. The most attractive possibility in this regard is gauge mediation. (For a review of gauge mediation and references to original work, see \GiudiceBP.) Since the SM gauge interactions are flavor blind, and the scale of SUSY-breaking is low, gauge mediation elegantly solves the SUSY flavor problem within a calculable framework.   But at the same time, the low scale of SUSY breaking implies that the gravitino is very light (anywhere from an eV to a GeV). Thus the MSSM superpartners are unstable to decay to the gravitino, and the direct connection between SUSY and dark matter would appear to be lost.\foot{Gravitino dark matter, while a logical possibility, is highly constrained and model dependent; see \GiudiceBP\ for a concise discussion of some of the issues. For simplicity, we will assume here that for whatever reason (low SUSY-breaking scale, late entropy production, low reheat temperature, etc.), the gravitinos do not comprise a significant fraction of the dark matter.}

The purpose of this paper is to explore a simple idea for how TeV scale dark matter could still arise naturally in a wide class of models of gauge mediation -- not from the MSSM, but from the SUSY-breaking sector. The class of models we will study are those based on generalized O'Raifeartaigh (O'R) models. These are renormalizable, weakly-coupled models of chiral superfields where SUSY is broken spontaneously by a tree-level F-term vev. They are probably the simplest and most straightforward examples of spontaneous SUSY breaking. They are also interesting and relevant for dynamical SUSY breaking, since they can arise as the low-energy effective description of SUSY gauge theories, as in \refs{\IntriligatorPU\IY-\IntriligatorDD}, or through retrofitting \DineGM. 

Generalized O'R models all share a key property -- the existence of ``pseudomoduli" -- that we will exploit to obtain dark matter. Pseudomoduli are chiral supermultiplets which are massless at tree-level, and whose scalar components extend to flat directions of the tree-level potential. They are lifted by radiative SUSY-breaking effects, and assuming this happens at one loop, the pseudomoduli masses will be (schematically)
\eqn\mpseudointro{\eqalign{
 & m_{scalar}^2 \sim {\alpha_h\over 4\pi} \left({F\over M}\right)^2\cr
  & M_{fermion} \sim {\alpha_h\over 4\pi}\left({F\over M}\right)
 }}
Here $\alpha_h=h^2/4\pi$ characterizes the coupling of the pseudomoduli to some fields of mass $M$ and SUSY splitting $F$ (we are assuming $\sqrt{F}\ll M$ for simplicity). If we identify these fields with the messengers of gauge mediation,  then in order to obtain a viable MSSM spectrum, we must have $F/M\sim 100$ TeV. Taking $h\sim 1$, we find that the scalar pseudomoduli, since they get one-loop mass-squareds, are typically $\sim$ 10 TeV. Meanwhile, the pseudomoduli fermions are similar to the gauginos in the MSSM and can naturally have TeV-scale masses. So the pseudomoduli spectrum takes the form
\eqn\mpseudointroii{\eqalign{
& m_{scalar}^2\sim (10\,\,{\rm TeV})^2\cr
&  M_{fermion} \sim 1\,\,{\rm TeV}
  }}
and we see that -- independent of any dark matter motivation -- this class of gauge mediation models more or less universally predicts the presence of TeV-scale particles in the spectrum, beyond those of the MSSM.
If these particles are stable (e.g.\ through a ${\Bbb Z}_2$ parity) and can annihilate efficiently to lighter states, then they can be viable dark matter candidates.\foot{It might also be interesting \ArvanitakiHQ\ to analyze the case where the dark matter can decay on cosmologically long timescales through higher dimension operators.}

Since the idea of pseudomoduli dark matter is completely general, it can easily accommodate different scenarios for dark matter annihilation. In particular, the pseudomoduli can be either charged or neutral under the SM gauge group. Of course, in the latter case, it must be able to annihilate to lighter fields that are neutral under the symmetry that keeps the pseudomoduli stable. The simplest possibility, which is motivated by recent cosmic ray anomales, is that the pseudomoduli  are charged under a light hidden sector gauge group. We will consider both SM-charged and SM-neutral pseudomoduli dark matter scenarios in this paper. Specifically, we will discuss in detail the case where the pseudomoduli are Higgsino-like (that is, transforming as $({\bf 2},+1/2)\oplus({\bf 2},-1/2)$ under $SU(2)\times U(1)_Y$), and the case where the pseudomoduli are SM-singlets but are charged under a $U(1)'$.

Our approach in this paper is meant to be ``bottom-up" and constructive. As such, it leaves many questions about the UV  unanswered.  In particular, the pseudomoduli in these examples are ``put in by hand" and their UV origin remains mysterious. Ideally, one would like to have an asymptotically-free SUSY gauge theory (e.g.\ some generalization of \IntriligatorDD) that dynamically generates these generalized O'R models with extra, stable pseudomoduli in the IR.\foot{Along these lines, the authors of \ZurZG\ analyzed the detailed phenomenology of a specific direct gauge mediation model \KitanoXG, and they suggested that the pseudomoduli fermions in this model could be cold dark matter candidates. However, the pseudomoduli in question were SM-singlets that could only annihilate through messenger loops, and moreover they were unstable to decay. So they do not fall under the ``standard WIMP paradigm" considered here.} This is an interesting problem that we leave for future work.

The outline of the paper is as follows. In section 2 we will describe the properties of generalized O'R models that lead to viable models of pseudomoduli dark matter. In section 3 we will discuss the possible quantum numbers of the pseudomoduli and review the various experimental constraints on two specific scenarios described above. In section 4 we will exhibit simple examples of pseudomoduli dark matter models that give rise to viable spectra. These serve to illustrate the general mechanism by which the pseudomoduli are stabilized and acquire masses.

\newsec{Pseudomoduli Dark Matter}

\subsec{General setup}

In this section we will describe in greater detail the general class of gauge mediation models in which the idea of pseudomoduli dark matter can be realized. Here we will not assume anything about the quantum numbers of the pseudomoduli. In the next section, we will specialize to more specific scenarios.

Pseudomoduli are automatic features of generalized O'Raifeartaigh models; as shown in \RayWK, the superpotential of such models can always be brought to the form
\eqn\WORgen{
W= f X + {1\over2}m_{ij}\phi_i\phi_j+{1\over2}\lambda_{ij}X\phi_i\phi_j+{1\over6}g_{ijk}\phi_i\phi_j\phi_k
}
with canonical K\"ahler potential. In this form, the tree-level scalar potential is extremized at $\phi=0$ with $X$ arbitrary. Thus $X$ is a flat direction of the tree-level potential which is present in every generalized O'R model \RayWK. In \KomargodskiJF, $X$ was referred to as the ``canonical pseudomodulus" and \WORgen\ as the ``canonical form" of the superpotential. 

Of course, the canonical pseudomodulus cannot be the dark matter, since (among other reasons) its fermionic component is the Goldstino, which is exactly massless in the absence of gravity. However, often in these models (e.g.\ \IntriligatorDD) there are many more pseudomoduli; sometimes these arise due to additional global symmetries.
We will denote these additional pseudomoduli by $Y$; they can be described by the superpotential interaction\foot{Actually, this is not the most general superpotential for models with additional pseudomoduli; see e.g.\ the magnetic quark pseudomoduli of \IntriligatorDD. But this ansatz certainly captures a large class, and we will restrict our attention to it for simplicity.}
\eqn\WORgenii{
\delta W = {1\over2}h_{ij}Y\phi_i\phi_j
}
Here we should point out that we are suppressing any gauge or flavor indices that $Y$ may carry; later on we will specialize to the case where $Y$ actually stands for a pair of pseudomoduli that are vector-like with respect to a gauge symmetry. We assume that the model respects an R-symmetry that forbids direct mass terms and other interactions for the pseudomoduli. Note that an R-symmetry is required for SUSY breaking on general grounds \NelsonNF, regardless of its origin. Ideally, it would arise as an accidental symmetry of the low-energy theory. 

The fields $Y$ can be made stable by imposing an appropriate symmetry. (Such a symmetry could also be an accidental consequence of the R-symmetry or global symmetries.) The simplest such symmetry, which we will focus on here, is a discrete ${\Bbb Z}_2$ symmetry under which $Y$ are charged. In that case, some subset of the $\phi$ fields must also be charged so that the interaction \WORgenii\ remains neutral. That is, $\phi$ should be split into two groups 
\eqn\phitoeta{
\phi\to \eta,\,\,\eta'
} 
with $\eta'$ charged under the ${\Bbb Z}_2$ and $\eta$ neutral. Then \WORgenii\ becomes
\eqn\WORgeniii{
W_{OR} = f X + {1\over2}(m_{ij}+\lambda_{ij}X)\eta_i\eta_j+{1\over2}(m_{ab}'+\lambda_{ab}'X)\eta_a'\eta_b'+h_{ia}Y\eta_i\eta_a'
}
In addition there could be cubic terms for the $\eta$ fields, but we ignore these since they do not contribute to the pseudomoduli dynamics at one loop.

Since the entire goal is to connect the dark matter mass scale with scale of SUSY breaking, some of the fields of the O'R model should be the messengers, as in \CheungES. This raises the attractive possibility that the ${\Bbb Z}_2$ symmetry that keeps the pseudomoduli stable could be simply an extension of the usual parity that forbids dangerous messenger-matter mixing. This would allow the messengers to decay and would solve the problem of otherwise stable messengers overclosing the universe \refs{\DineAG,\DimopoulosGY}.

Integrating out the fields that are massive at tree-level generates radiative masses for the pseudomoduli. This being the complete model, it is important that the dynamics of the O'R model also stabilize $X$ at a nonzero value, breaking the R-symmetry. We will assume that this happens at one loop, as in \ShihAV. Then generically masses for $Y$ are also generated at one-loop:
\eqn\Ymasses{
-\CL_{mass} = {1\over2}\left(m_{YY^*}^2|Y|^2+m_{YY}^2Y^2+M_{\psi_Y}\psi_Y^2 + c.c.\right)
}
In the appendix, we give general formulas for the masses appearing in \Ymasses, valid in the small $f$ limit. To have a consistent model,  the $Y$ scalars must be stabilized at the origin (otherwise the ${\Bbb Z}_2$ is broken and the dark matter can decay), and $\psi_Y$ should acquire its mass at one loop. In that case, as discussed in the introduction, $Y$ will naturally have a $\sim 10$ TeV mass, while $\psi_Y$ will naturally have a TeV scale mass.

To summarize, the requirements for a model of pseudomoduli dark matter are:

\lfm{1.} ${\Bbb Z}_2$ parity that makes the pseudomoduli stable.
\lfm{2.} R-symmetry that forbids direct tree-level pseudomoduli mass terms.
\lfm{3.} $X$ stabilized at nonzero value, breaking the R-symmetry.
\lfm{4.} $Y$ scalars stabilized at $Y=0$.
\lfm{5.} $\psi_Y$ mass generated at one-loop.
\lfm{6.} Tree-level interactions that allow the pseudomoduli to annihilate to lighter states. 

\subsec{Pseudomoduli cosmology}

Let us also take a moment to consider the cosmology of the $Y$ scalars. They behave very much like NLSPs in the early universe, in that they decay dominantly to $\psi_Y$ plus gravitino. (The main difference, of course, is that they are significantly heavier than usual NLSPs.) As long as the SUSY-breaking scale is sufficiently low, this decay happens rapidly, and the $Y$ scalars do not pose any cosmological problems. 

In more detail, the $Y$ lifetime is given by
\eqn\Ylifetime{
\tau_Y = {16\pi f^2\over m_Y^5}=\left({\sqrt{f}\over 100\,\,{\rm TeV}}\right)^4\left({10\,\,{\rm TeV}\over m_Y}\right)^5 \times 3\times 10^{-23}\,\,{\rm s}
}
{}From this, we learn that the decay $Y\to\psi_Y+\tilde G$ always happens before the time of BBN ($t\sim$ s), for all SUSY-breaking scales relevant for gauge mediation ($\sqrt{f}\lesssim 10^{7}$ TeV). 

One can also compute, using the relation $T\sim \tau^{-1/2}$, the temperature of the universe at the time of $Y$ decay:
\eqn\Ytemp{
T_Y \approx \left({m_{Y}\over 10\,\,{\rm TeV}}\right)^{5/2}\left({10^4\,\,{\rm TeV}\over \sqrt{f}}\right)^{2}\times 10\,\,{\rm GeV}
}
This should be compared with the temperature of $\psi_Y$ freeze-out, 
\eqn\Tfreeze{
T_{freeze}\sim M_{\psi_Y}/20\sim 50\,\,{\rm GeV}
} 
We see that the decay of $Y$ happens before $\psi_Y$ freeze-out, provided that 
\eqn\sqrtfbound{
\sqrt{f}\lesssim 10^4\,\,{\rm TeV}
} 
or equivalently, $m_{3/2}\lesssim 10$ keV. For simplicity, we will implicitly assume \sqrtfbound\ in the rest of the paper, so that the formula \relicdensii\ for the thermal relic density is valid.  While it would be interesting to consider the case where late decays of $Y$ scalars significantly affect the dark matter relic density, we will not do so here. 

\newsec{Quantum numbers}

As discussed in the introduction, in order to obtain the correct relic density for a TeV-scale dark matter particle, the annihilation cross section need only arise from an $\CO(1)$ coupling. This could either correspond to the electroweak interactions, or some additional interactions of the hidden sector.  In the first scenario, the dominant annihilation is to $W$ and $Z$ bosons. This is clearly the most economical and conservative scenario, but it is also the most experimentally constrained. In the second scenario, there are obviously fewer experimental constraints, so the possibilities are more varied here. The simplest possibility, which we will focus on here, is to assume that  the dark matter is a SM singlet but is charged under a hidden-sector $U(1)'$ gauge group. We assume that the $U(1)'$ is broken at a scale lighter than the dark matter mass and only interacts with the MSSM via kinetic mixing, as in the ``secluded" dark matter framework of \PospelovMP. (Recently, various cosmic ray anomalies have motivated the GeV scale as the scale of $U(1)'$ breaking \ArkaniHamedQN.) In this section, we will consider both scenarios in the context of pseudomoduli dark matter.

\subsec{SM-charged pseudomoduli}

Let us start by discussing the more conservative possibility. If we take the dark matter to be SM charged, then what are its quantum numbers? If we require perturbative $SU(5)$ unification, then we claim that the {\it only} possibility is that it transform as a ${\bf 5}\oplus{\bf\bar 5}$ of $SU(5)$, up to possible mixing with singlets. 

The argument is as follows: the dark matter must be TeV-scale and must come from complete $SU(5)$ multiplets, so the only GUT multiplets consistent with perturbative $SU(5)$ unification (and anomaly cancellation) are ${\bf 10}\oplus {\bf\bar{10}}$,  ${\bf 10}\oplus{\bf \bar 5}$, ${\bf\bar{10}}\oplus{\bf 5}$ or ${\bf 5}\oplus {\bf \bar 5}$. (We are not counting multiple copies of these representations as a separate case. Clearly, including these will not change the following argument.) The first is ruled out since it has no uncolored, neutral WIMP candidate. The second and third are essentially fourth generations, and the dark matter would have to be the neutrino- or sneutrino-like particle in the ${\bf \bar 5}$ or ${\bf 5}$, respectively. So it would have to acquire its mass via EWSB by mixing with a singlet. This setup is essentially ruled out by direct detection experiments, since the dark matter, being chiral, would have an unsuppressed elastic coupling to the $Z$.\foot{We thank N.~Arkani-Hamed for bringing this to our attention.} Furthermore, limits on fourth generation quarks from collider experiments are quite stringent \PDG, and preclude the possibility of perturbative Yukawa couplings up to the GUT scale. That leaves ${\bf 5}\oplus {\bf \bar 5}$ (and possibly additional singlets) as the unique possibility.  Note that our claim applies to any TeV-scale SM-charged dark matter beyond the MSSM, not just pseudomoduli dark matter. 

Motivated by this result, we will now specialize to the case of ${\bf 5}\oplus{\bf\bar 5}$ pseudomoduli dark matter and  discuss in detail the constraints on this scenario. In the previous section  we denoted the pseudomoduli by $Y$, neglecting any gauge indices; here it will obviously be preferable to change the notation slightly and denote the vector-like pseudomoduli by $Y$, $\tilde Y$.

Recall that under $SU(5)\to SU(3)\times SU(2)\times U(1)_Y$, a ${\bf 5}$ of $SU(5)$ decomposes into an $SU(3)$ triplet and an $SU(2)$ doublet:
\eqn\fivedec{
{\bf 5}\to\left({\bf 3},\,{\bf 1},\,-{1\over3}\right)\oplus \left({\bf 1},\,{\bf 2},\, {1\over2}\right)
}
By introducing supersymmetric splittings amongst the doublet and  triplet superpotential couplings in \WORgeniii, we can obviously make the colored pseudomoduli heavier than the uncolored ones. Indeed, this is motivated from unification -- if the doublet/triplet couplings unify at the GUT scale, then one expects $m_{3}\sim 2m_{2}$ at the weak scale from RG evolution. Moreover, in \ThomasWY\ it was shown that loops of electroweak gauge bosons always lift the electrically charged component above the neutral component by an amount $\CO(\alpha m_Z)$.
Thus the electrically neutral component of $SU(2)$ doublet pseudomoduli can naturally be the lightest particle in the GUT multiplet and can serve as a viable WIMP candidate.

Note that the triplets tend to be long-lived, since the same parity that makes the dark matter stable also makes the  triplets stable at the renormalizable level, but not necessarily at the level of higher-dimension (e.g.\ GUT-suppressed) operators. For a discussion of the interesting collider signatures and cosmology of long-lived, heavy, colored particles, see e.g.\ \refs{\ArvanitakiNQ,\ArkaniHamedQP}. Meanwhile, the charged components of the doublets could be very challenging to see at colliders \ThomasWY, since their decay products will tend to be extremely soft. Of course, if the doublets are as heavy as a TeV, then their direct production rate is negligible even at the LHC. But perhaps the triplets could help, if they are nearby in mass.

Calculations of the thermal relic density of $SU(2)$ doublet dark matter were performed in \MahbubaniPT. (See also \CirelliUQ, which has nice analytic formulas and generalizations to higher $SU(2)$ representations.) The upshot is that, in order to achieve the correct relic abundance \relicdens, the mass of the dark matter must be
\eqn\Msutwo{
M_{DM}\approx 1.1\,\,{\rm TeV}
} 

The precise details of the dark matter mass terms in the Lagrangian are important for direct detection experiments. These search for evidence of interactions between WIMPs and heavy nuclei via $Z$-exchange. In the $SU(2)\times U(1)$-symmetric limit, the only allowed mass term for the pair of doublets is
\eqn\massi{
\CL_{mass} = M_{\psi_Y}\psi_Y\tilde\psi_Y+c.c.
} 
so there are two degenerate mass eigenstates,
\eqn\twostates{
\psi_\pm = {\psi_{Y}\pm\psi_{\tilde Y}\over \sqrt{2}}
}
These are interchanged under their interaction with the $Z$ (i.e.\ there is no $\bar\psi_\pm\slashchar{Z}\psi_\pm $ vertex, only a $\bar\psi_\pm\slashchar{Z}\psi_\mp$ vertex), but since they are degenerate in mass, elastic scattering of the dark matter off the nucleus via $Z$ exchange is kinematically allowed. Consequently, there are rather strong bounds coming from direct detection experiments -- to the extent that TeV-scale dark matter of this type is already experimentally ruled out \EssigAZ. 

To avoid these constraints, one must introduce $SU(2)\times U(1)$-violating mass terms, 
\eqn\massii{
\delta\CL_{mass}=\delta M_{DM}\, \psi_Y^2+\delta \tilde M_{DM}\, \psi_{\tilde Y}^2+c.c.
} 
which split the mass eigenstates \twostates. If the splittings are large enough ($\gtrsim 10$ keV), nuclear scattering becomes inelastic and the direct detection bounds are correspondingly weakened (or avoided altogether). This idea of ``inelastic dark matter" has a long history; it was first proposed in the context of messenger dark matter by \HanWN. In \refs{\TuckerSmithHY\TuckerSmithJV-\ChangGD}, $\sim$ 100 keV splittings were proposed  as an elegant way of reconciling the DAMA/LIBRA anomaly with the null results of other direct detection experiments. However, very recently  \refs{\NussinovFT, \MenonQJ}\ it was argued that if the dominant annihilation is to $W$ and $Z$ bosons (as would be the case for $SU(2)$ doublets), the 100 keV splittings needed for DAMA would lead to enhanced capture of dark matter in the sun, and would  thus be inconsistent with bounds from observations of solar neutrino fluxes. Indeed, to satisfy these bounds, one requires $\delta M_{DM}\gtrsim$ MeV. So perhaps $SU(2)$ doublet dark matter cannot be consistent with DAMA.  Conversely, if the DAMA anomaly turns out to be correct after all, then the scenario of $SU(2)$ doublet dark matter will be disfavored.

We will not attempt to resolve all these experimental ambiguities; we will simply note that in this framework, the splittings \massii\ are free parameters which can arise from renormalizable couplings of the form
\eqn\deltaW{
\delta W = \lambda_i YH_d S_i + \tilde\lambda_i \tilde Y H_u S_i
}
where the $S$ fields are SM singlets (possibly from the O'R model \WORgeniii) that carry the same ${\Bbb Z}_2$ parity as the pseudomoduli. After integrating out the singlets at tree-level and substituting the Higgs vevs, we are left with operators of the form \massii, with the splittings related to the messenger scale by
\eqn\splittings{
\delta M_{DM} \sim {\lambda^2v^2\over M}
}
With $\lambda\sim \CO(1)$, this implies $M\sim  10^{5}$ or $10^4$ TeV in order to get the desired 100 keV or 1 MeV splittings. 

Finally, we should keep in mind that the requirement of perturbative gauge couplings up to the GUT scale may be too stringent. In particular, electric-magnetic duality could decrease the size of the $SU(5)$ representations in the UV, thereby averting an apparent Landau pole in the SM gauge couplings before the GUT scale (see e.g.\ \refs{\KitanoXG, \HabaRJ}\ for concrete examples of this phenomenon based on \IntriligatorDD).  So we are certainly not claiming to rule out the possibility of TeV-scale dark matter in higher $SU(5)$ representations. Indeed, it would be interesting to consider other $SU(5)$ representations, especially real representations such as the ${\bf 24}$. Such representations would naturally evade the current direct detection constraints even without any inelastic splittings \EssigAZ.

\subsec{SM-singlet pseudomoduli}

Now let us consider the possibility that the dark matter is a SM singlet but is charged (and vector-like) under a hidden-sector $U(1)'$ gauge group. Without loss of generality, we take the pseudomoduli to have charges $\pm1$ under the $U(1)'$. (We leave the analysis of non-abelian gauge groups for future work.) The annihilation cross section of $Y$ into on-shell dark photons (we ignore phase space threshold factors) is given by
\eqn\gdarkxsec{
\langle\sigma v\rangle ={\pi\alpha_d^2\over M_{DM}^2}\approx \left({\alpha_d\over 0.03}\right)^2 \left({1\,\,{\rm TeV}\over M_{DM}}\right)^2\times 3\times 10^{-26}\,\,\,{\rm cm}^3\,{\rm s}^{-1}
}
where $\alpha_d=g_d^2/4\pi$ and $g_d$ is the $U(1)'$ gauge coupling.

If the pseudomoduli are SM singlets, and they couple to the messengers of gauge mediation, then the cubic couplings in \WORgeniii\ imply that there must be ``link fields" present in the hidden sector, i.e.\ fields charged under both $SU(5)$ and $U(1)'$. The link fields will generate a kinetic mixing \HoldomAG\ between $U(1)'$ and $U(1)_Y$ given parametrically (up to finite logarithmic factors that depend on the link field masses) by
\eqn\kineticmixing{
\epsilon \sim {g'g_d\over 4\pi^2}
}
So in the absence of any surprise cancellations with other contributions to the kinetic mixing, and with $g_d\sim 1$, we expect $\epsilon\sim 10^{-4}-10^{-3}$ in these models.

Now, if the link fields are themselves messengers, i.e.\ they couple to the $X$ field in \WORgeniii, then there is standard gauge mediation to the dark sector, and we expect weak-scale soft masses in the dark sector just as in the MSSM. Then also by analogy with the MSSM, we would expect the scale of $U(1)'$ breaking to also be weak scale. This is a logical possibility not excluded by any experiment. But since we are motivated by recent cosmic ray anomalies to consider a scale of $U(1)'$ breaking much lower than the weak scale, we will not consider this case any further.

Instead, suppose the link fields are not messengers, i.e.\ they only have supersymmetric masses in \WORgeniii. This could be enforced by an appropriate R-symmetry. Then the leading mediation of SUSY breaking to the dark sector happens at {\it three} loops, via the pseudomoduli fields.
\eqn\mdarksq{\eqalign{
\delta m_{soft}^2&\sim q^2 \left({\alpha_d\over 4\pi}\right)^2\left({\alpha_h\over 4\pi}\right)\left({F\over M}\right)^2\cr
 }}
where $q$ is the $U(1)'$ charge of the dark sector scalar in question. If we rewrite \mdarksq\ using  \mpseudointro\ and \gdarkxsec, it becomes
\eqn\mdarksqii{\eqalign{
 \delta m_{soft}^2\sim q^2 \left({\langle\sigma v\rangle\over 3\times 10^{-26}\,\,{\rm cm}^3\,{\rm s}^{-1}}\right)\left({M_{DM}\over 1\,\,{\rm TeV}}\right)^3\left({F/M\over 100\,\,{\rm TeV}}\right)\times (20\,\, {\rm GeV})^2
}}
Keep in mind that this is a heuristic estimate and even the sign of \mdarksq\ is not certain. 
But it seems to indicate that there could be some tension between having $M_{DM}\gtrsim 1$ TeV and having GeV-scale soft masses in the dark sector, as suggested by various cosmic ray anomalies \ArkaniHamedQN. Of course, we did not actually calculate the loop diagrams that determine the dark-sector soft masses, and it is entirely possible that numerical factors could end up decreasing \mdarksqii. Also, having smaller charges in the dark sector helps. Clearly, it would be interesting to do a more detailed calculation in a specific model.

Finally, as discussed in \ArkaniHamedQP, these models can be tested by direct detection experiments, if the $U(1)'$ gauge boson is GeV-scale. However, it is always straightforward to evade these experiments by introducing splittings, as in the SM charged case. Also, these splittings could be valuable for explaining the DAMA/LIBRA anomaly, without the attendant problematic neutrino flux coming from solar WIMP capture   \refs{\NussinovFT, \MenonQJ}.

\newsec{Examples}

In this section, we will analyze two example models that illustrate the scenarios of Higgsino-like and $U(1)'$ pseudomoduli dark matter discussed in the previous section. But before describing these examples, let us first briefly review the features of a simple O'R model that breaks the R-symmetry spontaneously. This will serve as the basic building block from which we construct our examples of pseudomoduli dark matter in the following subsections. 

\subsec{An R-symmetry breaking O'R model}

The R-symmetry breaking O'R model is basically a vectorlike version of the model constructed in \ShihAV. It is characterized by the superpotential
\eqn\basicORW{
W_{OR} = f X+\lambda X(\eta_1\tilde\eta_2+\eta_2\tilde\eta_3)+m_1(\eta_1\tilde\eta_1+\eta_3\tilde\eta_3)+m_2\eta_2\tilde\eta_2
}
Here $\eta_i$, $\tilde\eta_i$ are taken to be singlets in this subsection. Later we will also take them to be charged under $SU(5)$, but this will not change the analysis.

This model respects a $U(1)_R$ symmetry under which the fields have R-charges
\eqn\RchargesbasicOR{
R(X)=2,\quad R(\eta_2)=R(\tilde\eta_2)=1,\quad R(\eta_1)=R(\tilde\eta_3)=-1,\quad R(\tilde\eta_1)=R(\eta_3)=3
}
Since there are fields with R-charges different from 0, 2, the R-symmetry can be broken spontaneously by the one-loop CW potential \ShihAV. In the limit of small SUSY breaking (which, as discussed in the introduction, we are assuming throughout for simplicity), the CW potential reduces to
\eqn\CWred{
V_{CW} = f^2 F\left ({\lambda X\over m_2},{m_1\over m_2}\right) + \CO(f^4)
}
So the dynamics of this model is controlled by a single dimensionless parameter 
\eqn\rdef{
r\equiv m_1/m_2
} 
and one finds that $X$ is stabilized at $\langle X \rangle\ne 0$ for $r<0.47$. A plot of $\langle X\rangle$ vs.\ $r$ is shown in figure 1. As $r\to 0$, the vev of $X$ approaches $\langle X\rangle\approx 0.25m_2/\lambda$. The maximum value of $\langle X\rangle$ is $\langle X\rangle\approx 0.33m_2/\lambda$.

\ifig\xminvsrfig{A plot of the vev of $X$  vs.\ the dimensionless parameter $r=m_1/m_2$ in the O'R model \basicORW.}{\epsfxsize=.75\hsize\epsfbox{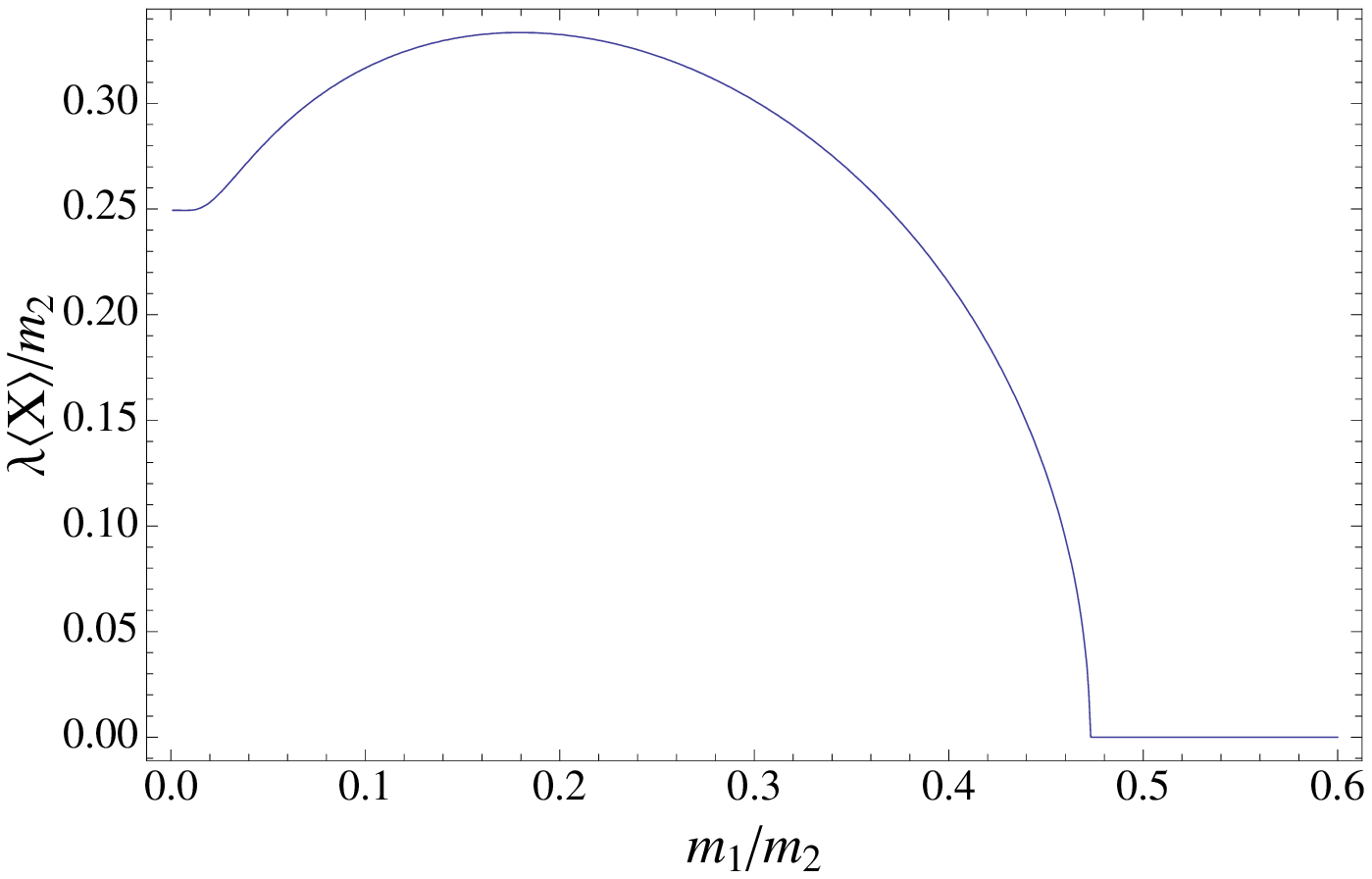}}

\subsec{An SM-charged example}

Now let us present a simple model of pseudomoduli dark matter that satisfies the requirements above and illustrates the main features of our idea. The model is constructed by starting with the R-symmetry breaking O'R model \basicORW, with all the fields taken to be singlets. Now we add to this model a ${\bf 5}\oplus{\bf\bar 5}$ pair of minimal gauge mediation messengers, denoted by $\eta'$, $\tilde\eta'$. Finally, we include a pair of pseudomoduli $Y$, $\tilde Y$ that also transform under ${\bf 5}\oplus{\bf \bar 5}$ of $SU(5)$. They couple to the messengers and to the singlets via cubic terms like in \WORgeniii. Thus the model we will study is:
\eqn\pmstoyex{
W =  \lambda'X \eta'\tilde\eta'+h (Y\tilde\eta'\tilde\eta_3+\tilde Y\eta'\eta_1)+W_{OR}(X,\eta,\tilde\eta)
}
For simplicity, we will assume here that all couplings are exactly $SU(5)$ symmetric; the various splittings required for consistency were discussed above and are trivial to implement here. 

This model respects an R-symmetry \RchargesbasicOR\ under which
\eqn\Rsymmetryi{
 R(\eta')=R(\tilde\eta')=0,\quad   R(Y)=R(\tilde Y)=3
 }
It also respects the requisite ${\Bbb Z}_2$ parity that keeps the pseudomoduli $Y$, $\tilde Y$ stable. The superpotential \pmstoyex\ is the most general renormalizable one consistent with the symmetries. In particular the R-symmetry forbids a direct mass term for the pseudomoduli. 

The dynamics of the model are controlled by two dimensionless parameters: $\lambda'$ and $r$ (which was defined in the previous subsection). For $\lambda'$ sufficiently small, $\lambda'\lesssim 0.45$, the potential for $X$ is dominated by the contribution from the O'R model \basicORW, and there can be an R-symmetry breaking minimum. Computing the masses of the fermions and scalars in $Y$, $\tilde Y$ using the formulas in the appendix, we find that there is a region of parameter space where the scalars are stabilized at the origin and the fermions have sizeable masses. A plot of this region is shown in figure 2; the contours shown are lines of constant $R$, where  the dimensionless quantity $R$ (not to be confused with the R-charges) is defined to be
\eqn\Rdef{
R \equiv \left({\alpha_h\over 4\pi}{f\over X} \right)^{-1} \times M_{\psi_Y} 
}
$R$ is a measure of the relative size between the dark matter mass and the MSSM soft masses, which are determined a la gauge mediation to be $m_{soft}\sim {\alpha_g\over 4\pi} {f\over X}$. To have a TeV-sized $\psi_Y$ mass with $\CO(1)$ couplings and a not-too-heavy superpartner spectrum, we would like to have $R$ not too much smaller than one.

\ifig\Xvspars{A contour plot of $R$ vs.\ the dimensionless parameters of the model \pmstoyex, $\lambda'$ and $r=m_1/m_2$. The regions where $Y$ are tachyonic and where $X$ has no R-symmetry breaking minimum are also indicated.}{\epsfxsize=.6\hsize\epsfbox{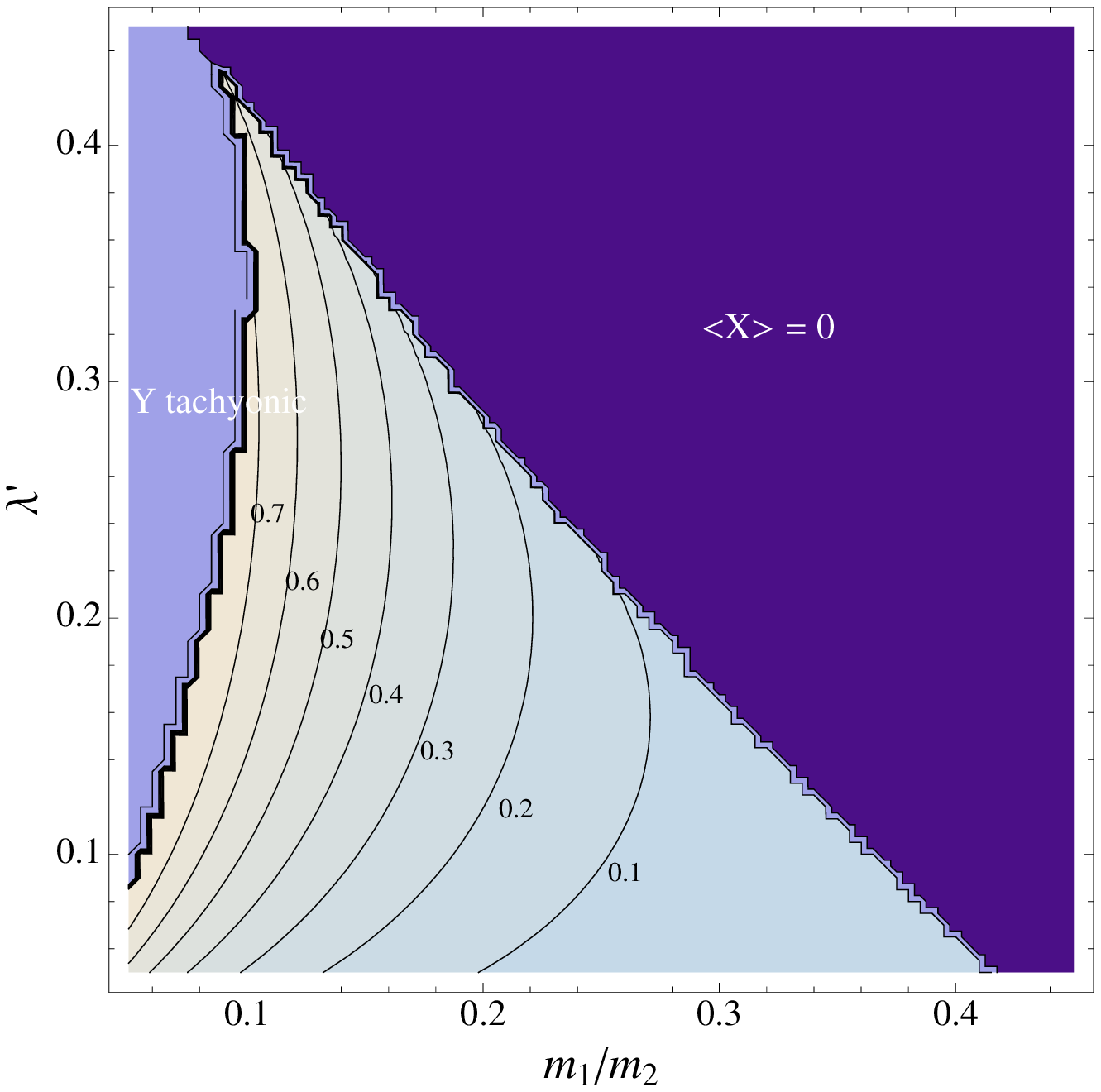}}

In fact, we see from fig.\ 2 that $R<1$ in all of the parameter space of this particular model. Nevertheless, there is a sizeable region of parameter space where it is large enough. For instance, if $R=0.5$, then taking $h= 1.1$ and $f/X=240$ TeV leads to an acceptable superpartner spectrum and $M_{\psi_Y}=1.1$ TeV.

\subsec{An SM-singlet example}

Next let us analyze an example of $U(1)'$ pseudomoduli dark matter. As discussed in the previous section, we take the pseudomoduli $Y$, $\tilde Y$ to have charges $\pm1$ under the $U(1)'$. The model we will consider starts again from the R-symmetry breaking O'R model \basicORW, but now with $\eta$, $\tilde\eta$ transforming as ${\bf 5}\oplus {\bf\bar 5}$'s of $SU(5)$. Introduce link fields $\eta'$, $\tilde\eta'$ transforming as $({\bf \bar 5},+1)\oplus ({\bf 5},-1)$ and with no direct coupling to $X$, and consider the superpotential:
\eqn\smsingletex{
W = m' \eta'\tilde\eta'+h(Y\tilde\eta'\tilde\eta_3+\tilde Y\eta'\eta_1)+W_{OR}(X,\eta,\tilde\eta)
}
The R-symmetry is now \RchargesbasicOR\ plus
\eqn\Rsymmetryi{
 R(\eta')=R(\tilde\eta')=1,\quad R(Y)=R(\tilde Y)=2
 }
This model is even simpler to analyze, since the dynamics of the $X$ field are now completely determined by the OR model, i.e.\ the $\eta'$, $\tilde\eta'$ fields do not contribute to the potential for $X$. Shown in fig.\ 3 is a plot of $R$ vs. $m'/m_2$ and $r$. We see that it is easy to achieve $\gtrsim$ TeV masses for the $Y$ fermions within the parameter space of the model. For example, taking $r=0.25$, $m'=0.5m_2$, one finds $R=1.4$. For $f/X=200$ TeV and $h=1$, this gives $M_{\psi_Y}=1.8$ TeV. Note that we are benefiting here from a factor of 5 enhancement coming from the fact that fundamentals of $SU(5)$ are running in the loop. This is a generic feature of SM singlet models, and it goes in the right direction, since the dark matter interpretation of the latest FERMI data seems to favor a heavier (multi-TeV) dark matter particle \MeadeIU.

Of course, this model has one obvious problem: there are no leading-order gaugino masses. That is, since $\det\CM=const$ in the messenger sector (the OR model \basicORW), it is a type I EOGM model \CheungES, and
\eqn\Mgvanishes{
M_{g} \sim f\partial_X\log\det\CM =0
}
This can be easily remedied by adding an additional MGM messenger.
 
 \bigskip
\ifig\myfermsinglet{A contour plot of $R$ vs.\ the dimensionless parameters of the model \smsingletex, $m'/m_2$ and $r=m_1/m_2$. The regions where $Y$ are tachyonic and where $X$ has no R-symmetry breaking minimum are also indicated.}{\epsfxsize=.6\hsize\epsfbox{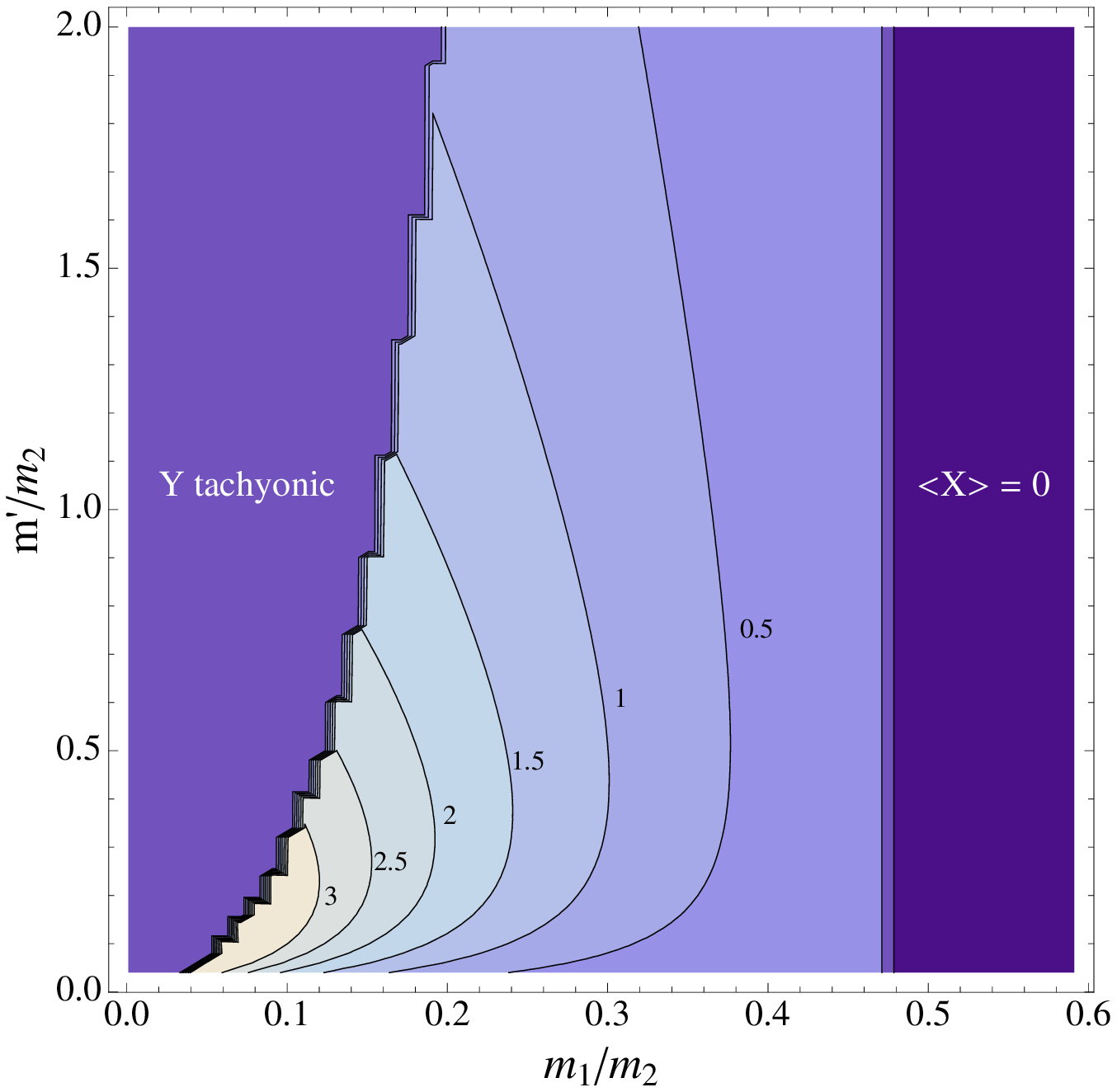}}

\bigskip
\bigskip

\noindent {\bf Acknowledgments:}

We would like to thank N.~Arkani-Hamed,  Z.~Komargodski, N.~Seiberg, L.~Senatore, S.~Thomas, N.~Toro and T.~Volansky for useful discussions. This work was supported in part by DOE grant DE-FG02-90ER40542.

\appendix{A}{General techniques for analyzing these models}

In the limit of small SUSY breaking, $f\ll m^2$, the theory below the messenger scale is described by a supersymmetric effective theory for the pseudomoduli. It takes the form
\eqn\Leff{
\CL_{eff} = \int d^4\theta\, K_{eff}(X,Y,\tilde Y)+\left(\int d^2\theta\, f X+c.c.\right)
}
 The effective one-loop K\"ahler potential in the theory below the messenger scale is given by the general formula:
\eqn\Keff{
K_{eff} = K_{can}-{1\over32\pi^2}\Tr\, \CM^\dagger \CM\log(\CM^\dagger \CM/\Lambda^2)
}
where $K_{can}$ is the canonical K\"ahler potential for all the pseudomoduli, and $\CM$ is the supersymmetric mass matrix of all the massive fields in the hidden sector (which is a function of all the pseudomoduli).

It is straightforward to use this effective theory to compute the effective potential for $X$ and the pseudomoduli spectrum. 
Imagine expanding out the effective K\"ahler potential in powers of $Y$, $\tilde Y$. The effective potential for $X$ comes about at zeroth order in $Y$, $\tilde Y$:
\eqn\VeffXex{
V_{eff}(X,X^*) = -f^2\partial_{X}\partial_{X^*}K_{eff}\Big|_{Y=\tilde Y=0}
}
We suppose that this is minimized for some $X=X_{0}$. The $Y$, $\tilde Y$ masses come at second order in the expansion:
\eqn\Ymassapp{
-\CL_{mass}\supset  m_{YY^*}^2 |Y|^2+m_{\tilde Y\tilde Y^*}^2|\tilde Y|^2 + (m_{Y\tilde Y}^2 Y\tilde Y + c.c. ) + \left( M_{\psi_Y}\psi_Y\psi_{\tilde Y} + c.c.\right)
}
where
\eqn\Keffmsq{\eqalign{
 & m_{AB}^2 = -f^2\partial_X\partial_{X^*}\partial_A\partial_{B}K_{eff}\Big|_{Y=\tilde Y=0,\,\,X=X_0}\cr
 & M_{\psi_Y} = f\partial_{X^*}\partial_Y\partial_{\tilde Y}K_{eff}\Big|_{Y=\tilde Y=0,\,\,X=X_0}
 }}
We will apply these formulas to calculate the dark matter spectrum in the example models of section 3.

\listrefs
\end